\documentclass[twoside,twocolumn,9pt]{article}
\usepackage{extsizes}
\usepackage[super,sort&compress,comma]{natbib} 
\usepackage[version=3]{mhchem}
\usepackage[left=1.5cm, right=1.5cm, top=1.785cm, bottom=2.0cm]{geometry}
\usepackage{balance}
\usepackage{widetext}
\usepackage{times,mathptmx}
\usepackage{sectsty}
\usepackage{graphicx} 
\usepackage{lastpage}
\usepackage[format=plain,justification=raggedright,singlelinecheck=false,font={stretch=1.125,small,sf},labelfont=bf,labelsep=space]{caption}
\usepackage{float}
\usepackage{fancyhdr}
\usepackage{fnpos}
\usepackage[english]{babel}
\usepackage{array}
\usepackage{droidsans}
\usepackage{charter}
\usepackage[T1]{fontenc}
\usepackage[usenames,dvipsnames]{xcolor}
\usepackage{setspace}
\usepackage[compact]{titlesec}
\usepackage{color}
\usepackage{soul}
\usepackage{booktabs}

\usepackage{epstopdf}

\definecolor{cream}{RGB}{222,217,201}

\begin{document}

\definecolor{JKcolor}{HTML}{00aa00}
\definecolor{JKcolorB}{HTML}{009900}
\definecolor{SPOTKANIEcolor}{HTML}{0044FF}
\definecolor{PJcolorB}{HTML}{FF1493}
\definecolor{JEScolor}{HTML}{000000}
\newcommand{\JES}[1]{\textcolor{JEScolor}{#1}}
\newcommand{\HIDE}[1]{{}}
\newcommand{\bigdif}[1]{{#1}}
\newcommand{\JK}[1]{\textcolor{JKcolor}{#1}}
\newcommand{\JKb}[1]{\textcolor{JKcolorB}{\ul{#1}}}
\newcommand{\JKrm}[1]{\textcolor{JKcolor}{\st{#1}}}
\newcommand{\JKcheckThis}[1]{\textcolor{JKcolor}{#1}}
\newcommand{\SPOTKANIE}[1]{\textcolor{SPOTKANIEcolor}{#1}}
\newcommand{\SPOTKANIEdone}[1]{\textcolor{SPOTKANIEcolor}{\st{DONE: #1}}}
\newcommand{\PJb}[1]{\textcolor{PJcolorB}{\ul{#1}}}

\pagestyle{fancy}
\thispagestyle{plain}
\fancypagestyle{plain}{

}

\makeFNbottom
\makeatletter
\renewcommand\LARGE{\@setfontsize\LARGE{15pt}{17}}
\renewcommand\Large{\@setfontsize\Large{12pt}{14}}
\renewcommand\large{\@setfontsize\large{10pt}{12}}
\renewcommand\footnotesize{\@setfontsize\footnotesize{7pt}{10}}
\makeatother

\renewcommand{\thefootnote}{\fnsymbol{footnote}}
\renewcommand\footnoterule{\vspace*{1pt}%
\color{cream}\hrule width 3.5in height 0.4pt \color{black}\vspace*{5pt}} 
\setcounter{secnumdepth}{5}

\makeatletter 
\renewcommand\@biblabel[1]{#1}            
\renewcommand\@makefntext[1]%
{\noindent\makebox[0pt][r]{\@thefnmark\,}#1}
\makeatother 
\renewcommand{\figurename}{\small{Fig.}~}
\sectionfont{\sffamily\Large}
\subsectionfont{\normalsize}
\subsubsectionfont{\bf}
\setstretch{1.125} 
\setlength{\skip\footins}{0.8cm}
\setlength{\footnotesep}{0.25cm}
\setlength{\jot}{10pt}
\titlespacing*{\section}{0pt}{4pt}{4pt}
\titlespacing*{\subsection}{0pt}{15pt}{1pt}

\fancyfoot{}
\fancyfoot[RO]{\footnotesize{\sffamily{1--\pageref{LastPage} ~\textbar  \hspace{2pt}\thepage}}}
\fancyfoot[LE]{\footnotesize{\sffamily{\thepage~\textbar\hspace{3.45cm} 1--\pageref{LastPage}}}}
\fancyhead{}
\renewcommand{\headrulewidth}{0pt} 
\renewcommand{\footrulewidth}{0pt}
\setlength{\arrayrulewidth}{1pt}
\setlength{\columnsep}{6.5mm}
\setlength\bibsep{1pt}

\makeatletter 
\newlength{\figrulesep} 
\setlength{\figrulesep}{0.5\textfloatsep} 

\newcommand{\topfigrule}{\vspace*{-1pt}%
\noindent{\color{cream}\rule[-\figrulesep]{\columnwidth}{1.5pt}} }

\newcommand{\botfigrule}{\vspace*{-2pt}%
\noindent{\color{cream}\rule[\figrulesep]{\columnwidth}{1.5pt}} }

\newcommand{\dblfigrule}{\vspace*{-1pt}%
\noindent{\color{cream}\rule[-\figrulesep]{\textwidth}{1.5pt}} }

\makeatother

\twocolumn[
  \begin{@twocolumnfalse}
\vspace{3cm}
\begin{tabular}{m{4.5cm} p{13.5cm} }

 & \noindent\LARGE{\textbf{Electronic structure and rovibrational predissociation of the $2^1\Pi$ state in KLi}} \\
\vspace{0.3cm} & \vspace{0.3cm} \\

 & \noindent\large{
 P. Jasik\textit{$^{a}$}, J. Kozicki \textit{$^{b}$}, T. Kilich\textit{$^{a}$}, J. E. Sienkiewicz \textit{$^{a\ast}$} and N. E. Henriksen\textit{$^{c}$}
 } \\

\vspace{0.3cm} & \vspace{0.3cm} \\

 & \noindent\normalsize{
Adiabatic potential energy curves of the $3^1\Sigma^+$, $3^3\Sigma^+$, $2^1\Pi$ and $2^3\Pi$ states correlating for large internuclear distance with the K(4s) + Li(2p) atomic asymptote were calculated. Very good agreement between the calculated and the  experimental curve of the $2^1\Pi$ state allowed for a reliable description of the dissociation process through a small ($\sim 20\ {\rm cm^{-1}}$ for $J=0$) potential energy barrier. 
The barrier supports several rovibrational quasi-bound states and explicit time evolution of these states via the time-dependent nuclear Schr\"odinger equation,
showed that the state populations decay exponentially in time.
We were able to precisely  describe the time-dependent dissociation process of several rovibrational levels and found that our calculated spectrum match very well with the assigned experimental spectrum. Moreover, our approach is able to predict the positions of previously unassigned lines, particularly in the case of their low intensity. 
} \\

\end{tabular}

\end{@twocolumnfalse} \vspace{0.6cm}

  ]

\renewcommand*\rmdefault{bch}\normalfont\upshape
\rmfamily
\section*{}
\vspace{-1cm}

\footnotetext{\textit{
$^{a}$~Faculty of Applied Physics and Mathematics, Gda\'{n}sk University of Technology, Narutowicza 11/12, 80-233 Gda\'{n}sk, Poland}}
\footnotetext{\textit{$^{b}$~Faculty of Civil and Environmental Engineering, Gda\'{n}sk University of Technology, Narutowicza 11/12, 80-233 Gda\'{n}sk, Poland}}
\footnotetext{\textit{$^{\ast}$~E-mail: "J.E. Sienkiewicz" <jozef.sienkiewicz@pg.edu.pl>}}
\footnotetext{\textit{$^{c}$~Department of Chemistry, Building 207, Technical
University of Denmark, DK-2800 Kgs. Lyngby, Denmark.}
}




\section{INTRODUCTION}\label{INTRODUCTION}

Studies of polar alkali dimers provide valuable insight into several basic phenomena, such as perturbations in excited states, potential curve crossings and avoided crossings, photodissociation,
photoassociation, and new quantum matter, namely  the Bose-Einstein condensate and ultracold two-fermionic species\cite{Moses2017}. The development  of
spectroscopic methods allows for very accurate measurements,
providing valuable data for the ground and excited states of these molecules.
From the theoretical point of view heteronuclear alkali metal dimers are
very attractive objects due to \JES{their} simple electronic structure and the possibility of treating
them as effective two-electron systems with separated
atomic cores. Pseudopotential methods with longtail
core polarization model potentials are well suited
to treat such systems\cite{Wiatr2017,Chaieb2017}.

 For years, the KLi dimer was an object of  spectroscopic experiments   
\cite{Walter1928,Weizel1930,Zmbov1977,Engelke1984,Sievers2015,Pazyuk2015} with the notable series of studies by the Warsaw group\cite{Jastrzebski2001,Grochola2002,Grochola2003,Salami2007,AdohiKrou2008,Jastrzebski2016} which provided, using Doppler-free polarization labeling technique, molecular constants and potential energy curves for the ground and several excited singlet states up to $11^1\Pi$ lying 33,000 $\rm cm^{-1}$ above the minimum of the ground state. Some potential energy curves, like in the case of the $1^3\Sigma^+$ state\cite{Salami2007}, were determined experimentally despite limited data sets of  vibrational energies.   The KLi dimer was also studied in the field of ultracold atomic and
molecular gases with strongly interacting two-fermionic species consisting of
$^{40}{\rm K}$ and $^6{\rm Li}$ atoms \cite{Wille2008, Voigt2009, Ridinger2011}.

Already in 1984,  M\"uller and Meyer\cite{Muller1984} performed extensive
all-electron SCF calculations on KLi along with \JES{a} careful treatment of intershell effects.  
In 1999,  Rousseau et al. \cite{Rousseau1999} performed CI calculations with nonempirical
one-electron pseudopotentials and appropriate
polarization potentials by means of the CIPSI program package\cite{Huron1973} obtaining 58 electronic
states of KLi.  Recently extensive theoretical ab initio studies were performed, including the electronic structure, transition dipole moments\cite{Dardouri2012}, and effect of inner-shell electrons on the molecular properties\cite{Xiao2013}. Previously we have calculated the low-lying potential energy curves, transition dipole moments and Franck-Condon factors in order to show effective schemes of  photoassociation reactions \cite{Miadowicz2013}. Also the electronic structure of KLi has been investigated by treating the K and Li atoms with the non-empirical relativistic effective core potentials \cite{Aldossary2014}. Very recently, the multireference coupled cluster method \JES{(MRCC)} has been
used to calculate 10 low-lying states of KLi\cite{Musial2016}. 

Important open questions concern  the exact shape of potential energy curves (PEC) of excited states.  The potential curve of the $2^1\Pi$ state derived from experiment  \cite{Grochola2003,Jastrzebski2009} exhibits a rotationless barrier. The position of the maximum and height of the barrier are decisive in precise time-dependent calculations of levels which lie above the dissociation limit.  

In our work,  we address the problem of  direct bound-to-free simulations, particularly  for quasi-bound rovibrational levels lying just below the potential barrier. This is the energy region in which the good agreement between semi-classical and fully quantum results is questionable.  We study the validity of the quantum bound-to-free approach near the potential barrier where quasi-bound levels are broad and their low intensity makes an additional challenge for experimentalists. 
We present the lines  of partial cross section for predissociation along with time-dependent populations. Initial quasi-bound wavepackets with precisely assigned  rotational and vibrational quantum numbers allows  us  to calculate the time-dependent population of rovibrational levels. 
To perform any time-dependent (dynamical) calculations it is preferable  to have potential energy curves of high accuracy, this is essential in description of the predissociation process. Thus, the scope of our work includes calculations of potential energy curves with the special focus on the $2^1\Pi$ state \JES{and getting new insight into the dynamics of  the rovibrational predissociation process. Details of this process showing the role of shapes of initial wavepackets  in forming the spectral lines are shown step by step on the molecular movies (see Supplementary material~\cite{Supplementary2018}).  To that end we show explicitly that the quasi-bound state populations decay exponentially in time. Moreover, our systematic study  of energies and widths of quasi-bond levels provides new valuable data for finding affective photoassociation schemes leading to Bose-Einstein condensates. Whenever is possible we made careful comparison with available experimental data to get better insight into the nature of the predissociation process. }



An overview of our computational method with the emphasis on the time-dependent method and boundary conditions is given in
the next section. In Section 3, we discuss results for the $2^1\Pi$ state and compare them with the  experimental data. Here also the spectrum and time-dependent population are presented and discussed. Conclusions are given in the last section. 

\quad
\section{THEORETICAL AND COMPUTATIONAL METHODS}\label{}

\subsection{Adiabatic potentials}


For each alkali atom only the valence electron is treated explicitly. This approach with two valence electrons for the whole alkali molecule \JES{has been already described in our earlier papers} and is proved to give reliable results, particularly for excited states \citep{Jasik2018,Wiatr2017,Wiatr2015,Szczepkowski2013,Lobacz2013,Jasik2007,Jasik2006}. Also, it can be applied for bigger molecules and even clusters, since the dimension of \JES{the} active space and \JES{the} number of configurations which has to be taken into account are relatively small. 

Here the core electrons of K atoms are represented by the pseudopotential
ECP18SDF \citep{Fuentealba1982}. We use the basis sets for potassium which come with ECP18SDF \citep{Fuentealba1988} for $s$ and $p$ functions and ECP10MDF pseudopotential \citep{Lim2005} for $d$ and $f$ functions. Additionally, these basis sets are augmented by seventeen $s$ functions, six $p$ functions, ten $d$ functions and six $f$ functions. The exponential coefficients of the uncontracted Gaussian Type Orbitals (GTO) \JES{are listed in Supplementary material~\cite{Supplementary2018}}.

The core electrons of the Li atom are represented by the pseudopotential ECP2SDF \cite{Fuentealba1982}. The basis set for the $s$ and $p$ orbitals, which comes with this pseudopotential \citep{Fuentealba1988} is enlarged by functions for $d$ and $f$ orbitals given by Prascher {\it et al.} \citep{Prascher2011} and assigned by cc-pV5Z. Additionally, our basis set was augmented by six $s$ short range correlation functions, seven $p$ functions, seven $d$ functions and five $f$ functions. Also, we added to the basis the following set of diffuse functions: three $s$ functions, four $p$ functions, three $d$ functions and three $f$ functions. Coefficients of the exponents of the primitive Gaussian orbitals \JES{are listed in Supplementary material~\cite{Supplementary2018}}.

\JES{In our calculations of core polarization potentials, the static dipole polarizabilities of the atomic cores \cite{Fuentealba1982} are taken as 5.354 and 0.1915 a$_0^3$ for K and Li, respectively. In turn the cut-off parameters are equal to 0.296519 a$_0^{-2}$ for K and 0.8351 a$_0^{-2}$ for Li.}  

To calculate adiabatic potential energy curves of the KLi molecule we use the  multiconfigurational self-consistent field/complete active space self-consistent field (MCSCF/CASSCF) method and the multi-reference singles and doubles configuration interaction (MRCISD) method. All calculations are performed by means of the MOLPRO program package \cite{MOLPRO2006}. We obtained adiabatic potential energy curves for four excited states, namely $^{1,3}\Sigma^+$ and $^{1,3}\Pi$ correlating to the K(4s)+Li(2p) atomic asymptote. 


\subsection{Molecular quantum dynamics}

The time-dependent approach which is mathematically equivalent to the time-independent one can be regarded as a complementary tool and is often used in studying photodissociation processes.
Here, it serves as an alternative and quite illustrative method for testing results of our structural calculations.

We start our consideration from the time-dependent Schr\"{o}dinger equation written in the following form

\begin{equation}\label{TDSE}
\imath \hbar \frac{\partial}{\partial t} \Phi(R,t) = H_{J}^{nuc} \Phi(R,t),
\end{equation}
where $\Phi(R,t)$ is the time-dependent wavepacket moving on the effective potential energy curve $U_J(R)$ 
and the nuclear Hamiltonian \JES{is taken as  
$H_{J}^{nuc} = -\frac{\hbar^2}{2\mu}\frac{\partial^2}{\partial R^2} + U_J(R).$ }


The evolving wavepacket $\Phi(R,t)$ is a solution of Eq.~(\ref{TDSE}).
We choose as initial state, $\Phi(R,t=0)=\Psi_{E,J}(R)$, a quasi-bound state which can be calculated very accurately by the methods implemented in
the program LEVEL~\cite{LeRoy2011}. \JES{This approach allows to calculate the population for the particular state labeled by $(v,J)$.}
The wavepacket is tunneling away from its starting position with the main amplitude
located inside the potential energy barrier. 
The time-dependent population $P(t)$ is calculated in the range from $R=0$ till $R_{max}$, 
where $R_{max}$ is chosen sufficiently to the right of the outermost classical turning point such that $P(t=0)=1$.
The population is calculated as

\begin{equation}\label{population_J}
P(t) =  \int_0^{R_{max}} | \Phi({R}, t) |^2\, dR.
\end{equation}

The process is also described by the time-dependent autocorrelation function

\JES{\begin{equation}\label{funkcja autokorelacji}
S(t) =  \int \Phi^\star({R}, t=0) \,  \, \Phi({R}, t) \, dR.
\end{equation}
}
In our case the autocorrelation function describes the evolution of the initial nuclear eigenfunction in the excited electronic state. 

We determine the spectrum by the inverse Fourier transform of
$S(t)$\JES{~\cite{Schinke1993}} as follows
\JES{\begin{equation}\label{tacs}
\sigma(E(\nu,J)) 
= \int_{-\infty}^{\infty} \mathrm{e}^{\imath E(\nu, J) t / \hbar} \, S(t) \, dt.
\end{equation}
}
\JES{In our calculations the above integral is estimated over the range <0,T> with fast Fourier transform (FFT) routines~\cite{2005Frigo}.}


A computer code for calculating the quantum dynamics~\cite{Kosloff1984,Kosloff1997,Kosloff1988}  was implemented in a computational framework developed by one of the authors~\cite{Kozicki2008,Kozicki2009}.
In our approach the Chebyshev polynomial recurrence relation formula~\cite{Kosloff1984,Abramowitz2013} was modified by multiplying it by the absorbing boundary conditions term $e^{-\gamma(R)}$~\cite{Mandelshtam1995a,Mandelshtam1995b}. The resulting absorbing potential acts as the commonly used imaginary absorbing potential, such that the calculations are numerically stable in the Kosloff calculation method. 

In the calculations of the autocorrelation function (Eq.~\ref
{funkcja autokorelacji}), the propagation time is 1~ns which is sufficient for estimation of the integral in Eq.~(\ref{tacs}). In Eq.~(\ref{population_J}), we set the value of $R_{max}$ to be equal to 57~a$_0$ (30~\AA). There are 8192 points in the whole grid and 4669 points in the integration grid (since integration excludes the region of the absorbing boundary conditions). In order to avoid the interference between the outgoing and incoming waves on the periodic grid an absorbing potential is placed in the range from 57 to 100 a$_0$ (30~\AA{} to 53~\AA). This potential smoothly absorbs the wavepacket~\cite{Mandelshtam1995a}.

\begin{figure}
\centering
\includegraphics[height=\columnwidth,angle=270]{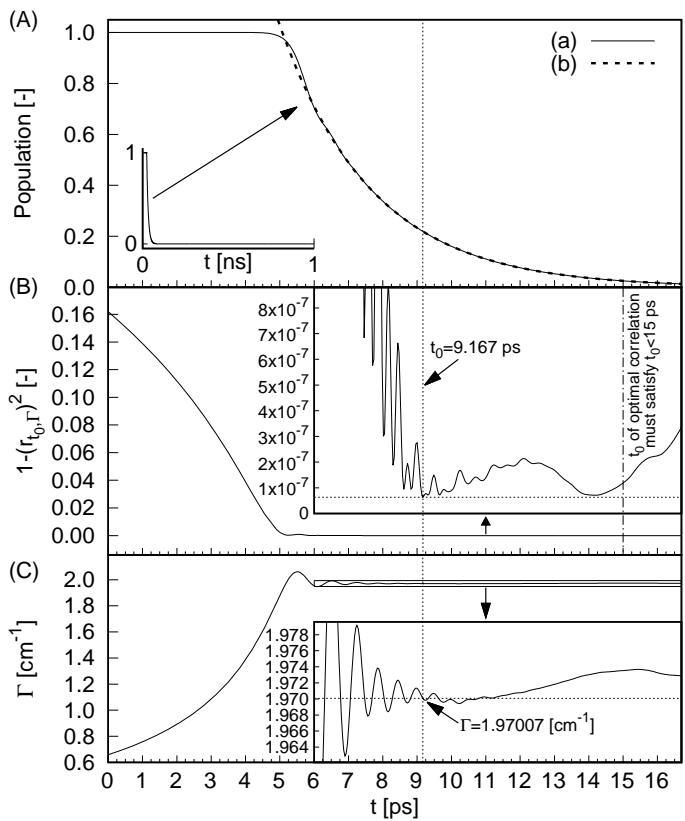}
\caption{{The procedure of finding optimal FWHM $\Gamma$ [cm$^{-1}$] shown for the level $J=37$ and $v=16$ of the 2$^1\Pi$ state. (Aa) Population of time--evolving  wavepacket is calculated up to 1~ns, but only the first 17~ps are shown here for clarity. (Ab) The obtained best fit population decay function $e^{-t/\tau}$, the population values to the left of $t_0$ were not used in the fit. (B) The value of Pearson's correlation coefficient $r_{t_0,\Gamma}$ (plotted as $1-(r_{t_0,\Gamma})^2$) as a function of time $t_0$ where the least squares fitting starts. (C) The value of FWHM $\Gamma$ [cm$^{-1}$] obtained from least squares fitting starting at each $t_0$.}}
\label{fig:R2_example}
\end{figure}

The obtained population curve $P(t)$ is roughly constant in  the first 5~ps (see Fig.~\ref{fig:R2_example}A), which is the time for the wavepacket to reach the $R_{max}$ point. After this the population (Eq.~\ref{population_J}) starts to decay. 
We observe that the population follows an exponential decay $e^{-t/\tau}$, with a decay constant $\tau$.
Here we demonstrate the exponential decay of population based on explicit time-propagation of a quasi-bound state. Often the
exponential decay of such states is described by introducing an imaginary part in the state energies~\cite{Schinke1993,Henriksen2008}. 
An exponential decay in population leads by using  Eq.~\ref{tacs} to Lorentzian line shapes with FWHM equal to $\Gamma=\hbar/\tau$~\cite{Schinke1993,Henriksen2008}. 

The time $t_0$ for best determination of the decay constant  $\tau$ (and correspondingly $\Gamma$) was determined by performing a parametric least squares fitting, where the $t_0$ was adjusted in order to maximize the Pearson's correlation coefficient $r_{t_0,\Gamma}$ value of the fit (see Fig.~\ref{fig:R2_example}B). The dependence of $\Gamma$ on $t_0$ (at which the least squares fitting starts) is shown on Fig.~\ref{fig:R2_example}C. It can be seen that a "perfect" fit of $\Gamma$ is found in a very narrow range (see inset of Fig.~\ref{fig:R2_example}C) in the order of 0.02 cm$^{-1}$. When scanning the entire time range $t_0\in$[5~ps,1~ns], the least squares fitting had a tendency to produce "perfect" fits $r_{t_0,\Gamma}=1.0$ at very large $t_0$ (of the order of 0.1~ns), hence to force inclusion of the beginning of the population decay (which starts around 5~ps) the fitting was constrained to start before $t_0<15$~ps.



\quad
\section{RESULTS AND DISCUSSION}
\subsection{Born-Oppenheimer potentials}

\begin{figure}
\centering
\includegraphics[height=7cm]{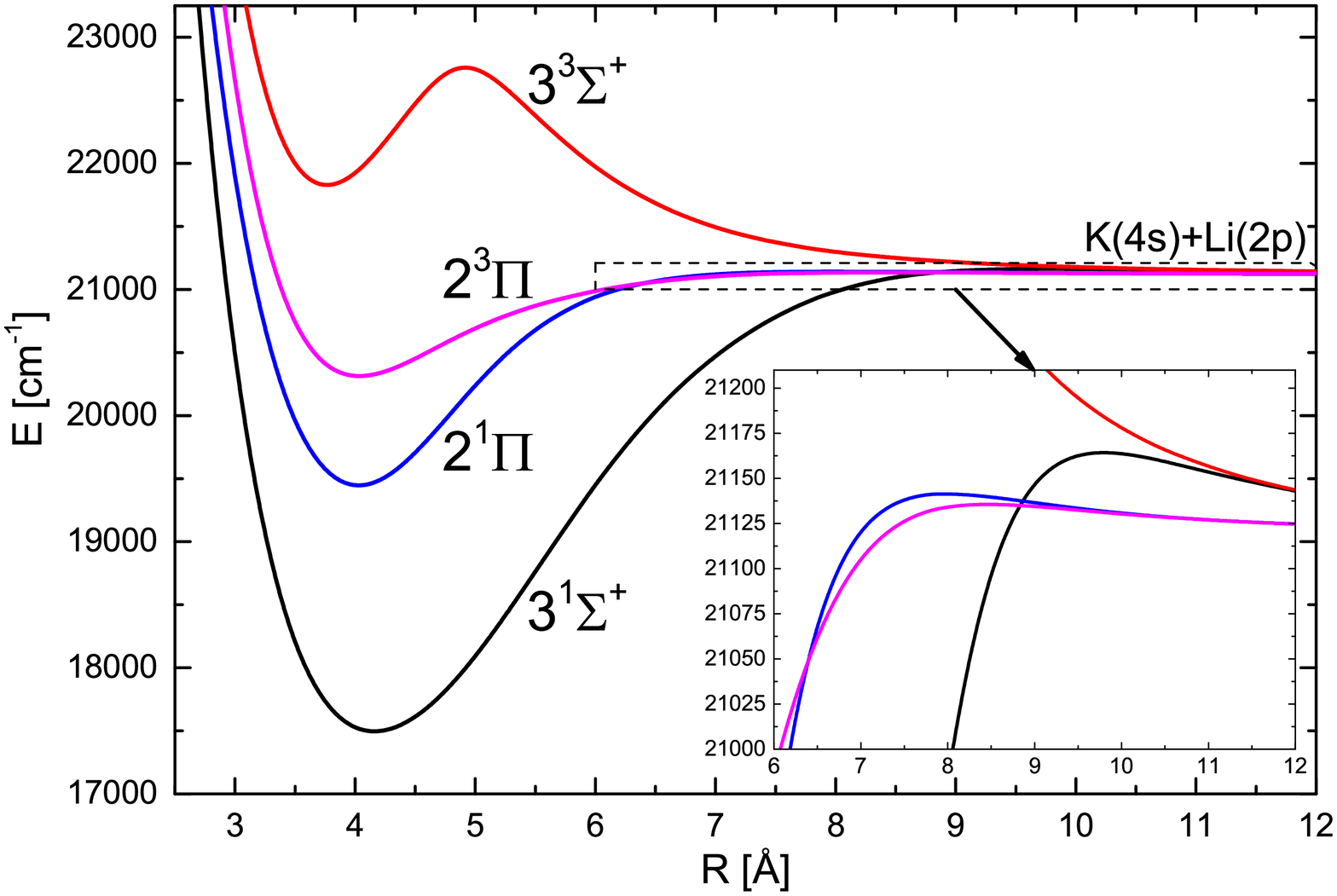}
\caption{Adiabatic potential energy curves of KLi correlating with the K(4s) + Li(2p) atomic asymptote. 
}
\label{fig:other_potentials}
\end{figure}

In order to obtain the broader picture besides the potential energy curve of the $2^1 \Pi$ state, we present in Fig.~\ref{fig:other_potentials} three other states correlating with the same atomic asymptote K(4s) + Li(2p). These are $3^1 \Sigma^+$, $2^3 \Pi$ and $3^1 \Sigma^+$. A table with numerical values are available in Supplementary material~\cite{Supplementary2018}. The inset in Fig.~\ref{fig:other_potentials} gives a closer view of the potential barriers, including the quite distinct one of the  $2^1 \Pi$ state, which plays an important role in the proper interpretation of a part of spectrum. Comparison between spectroscopic parameters calculated from our potential curves and those provided by experiment and other calculations are given in Tab. 1. The shape of our $3^1 \Sigma^+$ potential curve is very close to the most recent experimental one \cite{Grochola2002}, since the differences between respective parameters describing well depths D$_e$ and harmonic frequencies  $\omega_e$ are very small. Our calculated well minimum position R$_e$ lies very close to the experimental ones \cite{Grochola2002,Pashov1998}, although the quite recent theoretical result of Daudouri et al. \cite{Dardouri2012} lies even closer to the experimental values.
There is no experimental data for the $3^3 \Sigma^+$ state \JES{which has an} exotic shape. Its well minimum lies on the energy scale above the dissociation  limit and the well is adequately deep to support some quasi-bound levels with predicted $\omega_e$ around 152 cm$^{-1}$. 

\begin{figure}
\centering
\includegraphics[height=7cm]{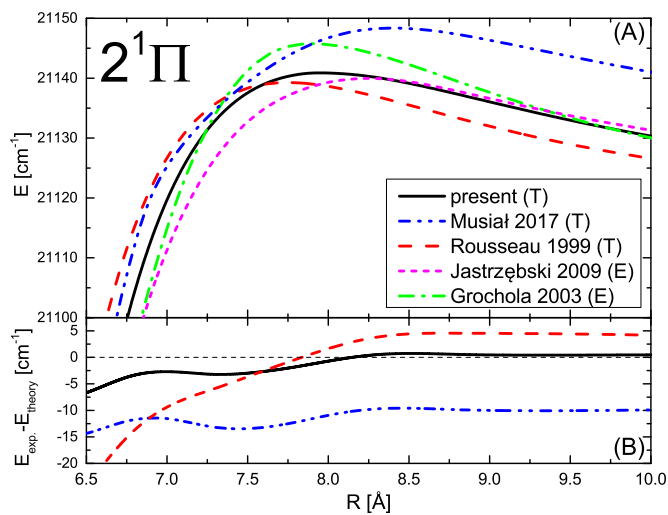}
\caption{(A) The potential energy barrier in the 2$^1\Pi$ state. Present theoretical results (black solid line) are compared with results derived from experimental data by Jastrz\c{e}bski et al. \cite{Jastrzebski2009} (magenta short-dash line) and Grochola et al. \cite{Grochola2003} (green dash-dot line) as well as with the other recent theoretical results given by Musia\l{} et al. \cite{Musial2017} (blue dash-dot-dot line) and Rousseau et al. \cite{Rousseau1999} (red-dash line). (B) Differences in energy between the experimentally determined potential \cite{Jastrzebski2009} of the 2$^1\Pi$ state and the present PEC as well as the other theoretical ones \cite{Musial2017,Rousseau1999}.}
\label{fig:potential_barrier}
\end{figure}

\begin{figure}
\centering
\includegraphics[height=7cm]{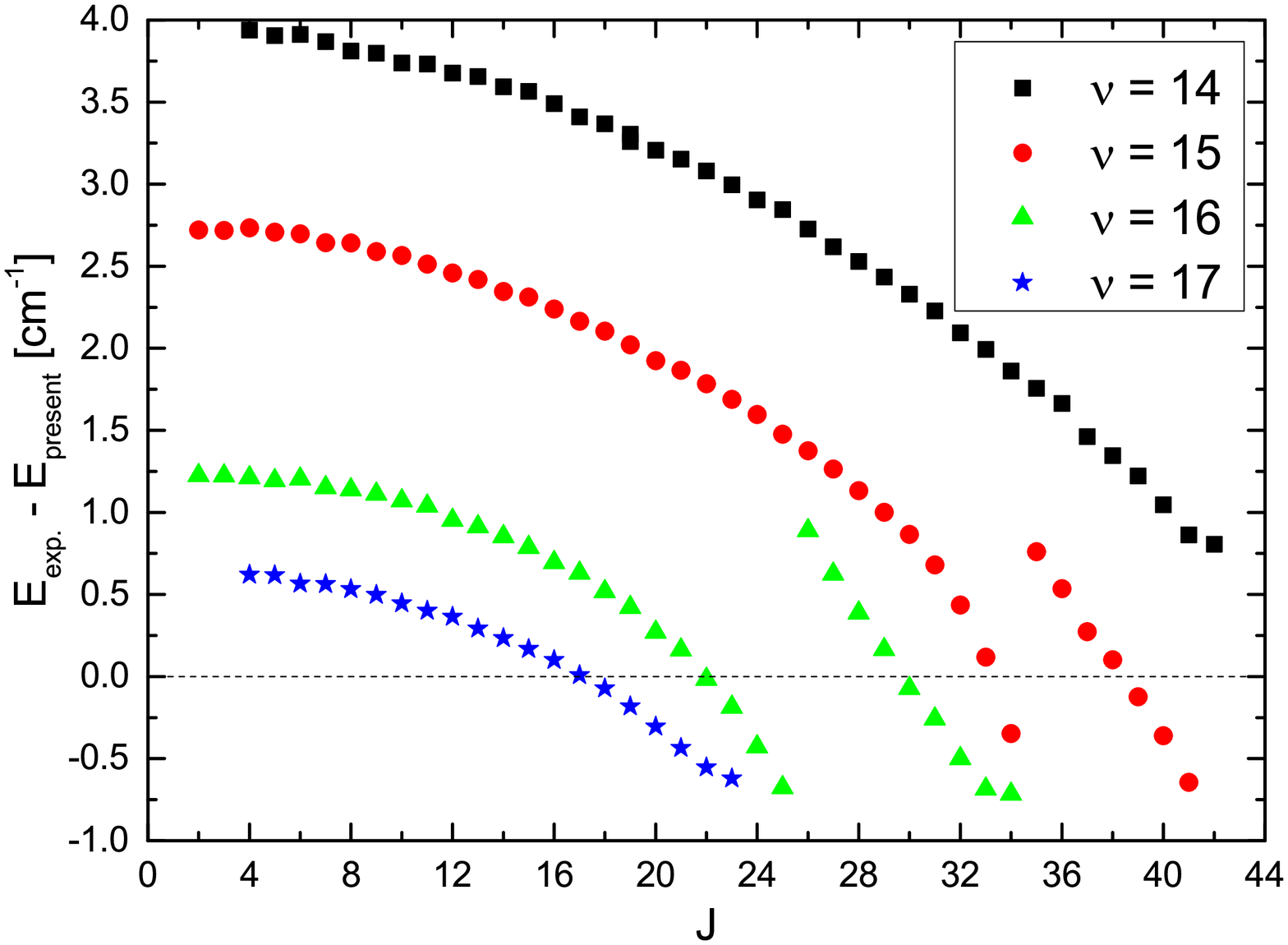}
\caption{Differences between experimental \cite{Jastrzebski2009} and our calculated  term values in the $v = 14$, $15$, $16$ and $17$ levels of the 2$^1\Pi$ state. }
\label{fig:term_differences}
\end{figure}

In case of the $2^1 \Pi$ state, the comparison of our spectroscopic parameters  R$_e$, D$_e$, $\omega_e$ and B$_e$ gives a very good consistent agreement with experimental data of Grochola et al. \cite{Grochola2002}.  It means that the description of the shape is very reliable. We also notice that two parameters R$_e$ and D$_e$ of the newest calculations of Musia{\l} et al.~\cite{Musial2017} are even closer to experimental ones. For this state the precise theoretical description of the rotationless potential barrier is very important due to the existence of refined experimental data by Jastrz\c ebski et al. \cite{Jastrzebski2009}. The position of the barrier's maximum and its height for two available sets of experimental data and three chosen theoretical results are given in Tab. 2. In turn, Fig.~\ref{fig:potential_barrier} shows the comparison between experimental data and three theoretical results of the barrier shape. Appropriate differences (Fig.~\ref{fig:potential_barrier}B) clearly show the almost perfect match between our potential barrier and the experimental one. The difference curve of Musia{\l} et al.~\cite{Musial2017} shows systematic discrepancy of 15 cm$^{-1}$ while the curve of Rousseau et al. \cite{Rousseau1999} along with increase of internuclear distance shows substantial improvement and  reasonable stabilization after crossing with the reference 0-line. The differences between experimental \cite{Jastrzebski2009} and our theoretical term values of the levels supported by the potential barrier (Fig.~\ref{fig:term_differences}) are rather small, not exceeding 4 cm$^{-1}$ in the $v= 14$ level with the mean absolute error (MAE) equal to 2.78 cm$^{-1}$ and only 0.5 cm$^{-1}$ in the  $v = 17$ level with MAE=0.38 cm$^{-1}$.  In the case of $v = 15$ and 16 MAE equals to 1.62 and 0.71, respectively. The levels $v=15$ and 16 exhibit strong perturbation with the visible shifts caused by interaction with the nearby $3^1 \Sigma^+$ and $2^3 \Pi$ states. 

For the other triplet $\Pi$ state there is no experimental data (Tab. 1). The comparison among theoretical results shows consistency between our R$_e$, D$_e$ and $\omega_e$ parameters and those of Musia{\l} et al.

\begin{table*}[ht]
  \caption{Spectroscopic parameters R$_e$ [\AA], D$_e$, $\omega_e$, B$_e$, and T$_e$ [cm$^{-1}$] for the excited states correlated in the atomic limit to the Li(2p) + K(4s) asymptote.\\[0.3cm]}
  \label{tbl:parameters}
\small
\begin{tabular*}{\textwidth}
  {@{\extracolsep{\fill}}l l c c c c c}
  \hline
State & Author & R$_e$ & D$_e$ & $\omega_e$ & B$_e$ & T$_e$ \\
  \hline
 &  &  &  &  &  & \\
$3^1 \Sigma^+$ & present (theory) & 4.158 & 3629 & 114.91 & 0.1628 & 17639 \\
 & Grochola 2002 (exp.) \cite{Grochola2002} & 4.192 & 3619 & 115.41 & 0.1614 & 17501 \\ 
 & Pashov 1998 (exp.) \cite{Pashov1998} & 4.190 & 3619 & 115.41 & 0.1615 & 17501 \\
 & {Musia\l{} 2017 (theory)} \cite{Musial2017} & 4.169 & 3597 & 116.32 &  & 17452 \\
 & Musia\l{} 2016 (theory) \cite{Musial2016} & 4.152 & 3369 & 110.92 &  & 17669 \\
 & Al-dossary 2014 (theory) \cite{Aldossary2014} & 4.154 & 3345 &  &  & 17699 \\
 & Dardouri 2012 (theory) \cite{Dardouri2012} & 4.186 & 3554 & 115.20 &  & 17767 \\
 & Rousseau 1999 (theory) \cite{Rousseau1999} & 4.128 & 3477 & 114.27 & 0.1649 & 17647 \\
 &  &  &  &  &  &  \\
$3^3 \Sigma^+$ & present (theory) & 3.762 & -700 & 152.13 & 0.1980 & 22587 \\
 & {Musia\l{} 2017 (theory)} \cite{Musial2017} & 3.783 & -774 & 152.92 &  & 21823 \\
 & Musia\l{} 2016 (theory) \cite{Musial2016} & 3.770 & -876 & 163.93 &  & 21914 \\
 & Al-dossary 2014 (theory) \cite{Aldossary2014} & 3.837 & -836 &  &  & 23865 \\
 & Dardouri 2012 (theory) \cite{Dardouri2012} & 3.773 & -918 &  &  & 22616 \\
 &  &  &  &  &  & \\
$2^1 \Pi$ & present (theory) & 4.028 & 1679 & 128.75 & 0.1737 & 19589 \\
 & Grochola 2003 (exp.) \cite{Grochola2003} & 4.043 & 1664 & 128.98 & 0.1735 & 19456 \\
 & Pashov 1998 (exp.) \cite{Pashov1998} & 3.713 & 1686 & 135.84 & 0.2056 & 17573 \\
 & {Musia\l{} 2017 (theory)} \cite{Musial2017} & 4.033 & 1664 & 132.62 &  & 19385 \\
 & Musia\l{} 2016 (theory) \cite{Musial2016} & 4.000 & 1682 & 134.11 &  & 19356 \\
 & Al-dossary 2014 (theory) \cite{Aldossary2014} & 3.990 & 1438 &  &  & 20155 \\
 & Dardouri 2012 (theory) \cite{Dardouri2012} & 4.069 & 1300 & 120.92 &  & 20410 \\
 & Rousseau 1999 (theory) \cite{Rousseau1999} & 4.022 & 1584 & 132.30 & 0.1765 & 19541 \\
 &  &  &  &  &  & \\
$2^3 \Pi$ & present (theory) & 4.030 & 811 & 103.65 & 0.1701 & 20457 \\
 & {Musia\l{} 2017 (theory)} \cite{Musial2017} & 4.078 & 803 & 103.88 &  & 20246 \\
 & Musia\l{} 2016 (theory) \cite{Musial2016} & 4.109 & 674 & 96.49 &  & 20365 \\
 & Al-dossary 2014 (theory) \cite{Aldossary2014} & 4.032 & 688 &  &  & 20458 \\
 & Dardouri 2012 (theory) \cite{Dardouri2012} & 4.011 & 612 & 123.60 &  & 21090 \\
 & Rousseau 1999 (theory) \cite{Rousseau1999} & 4.022 & 736 & 108.90 & 0.1732 & 20388 \\
 &  &  &  &  &  & \\
\hline
\end{tabular*}
\end{table*}

\begin{table*}[ht]
  \caption{Comparison of the barrier position R$_{bar}$ [\AA] and barrier height $\Delta E_{bar}$ [cm$^{-1}$] relative to the dissociation limit of the 2$^1\Pi$ state with other theoretical and experimental results.\\[0.3cm]}
  \label{tbl:barriers}
\small
\begin{tabular*}{\textwidth}
  {@{\extracolsep{\fill}}c c c c c c}
  \hline
  & This work & Exp.\cite{Jastrzebski2009} & Exp.\cite{Grochola2003} & {Theory \JES{(MRCC)}}\cite{Musial2017} & Theory \JES{(pseudopotentials+CI)}\cite{Rousseau1999} \\
\hline
 R$_{bar}$ & 7.98 & 8.20 & 7.79 & 7.93 & 7.62 \\
 $\Delta E_{bar}$ & 20.73 & 20.80 & 26.50 & 20.35 & 27.50 \\
\hline
\end{tabular*}
\end{table*}

\begin{table*}[ht]
\caption{Energies and widths of chosen quasi-bound levels from the 2$^1\Pi$ state. Present results obtained by time-dependent calculations (Present,TDSE) and by means of the program LEVEL (Present) are compared with those measured in the experiment~\cite{Jastrzebski2009}. Only levels with the f rotational symmetry are listed. The double column $\Delta E$ gives differences between Experiment and respectively Present,TDSE and Present. $\Delta \Gamma$ gives the relative difference between level widths calculated as (Present - Present,TDSE)/Present,TDSE.}
\label{tbl:comparison_all}
      \begin{tabular*}{\textwidth}{@{\extracolsep{\fill}}r l|c c c c|l l r}
      	\toprule
      ~   & ~  &                     \multicolumn{4}{c|}{Energy [$cm^{-1}$]}                                & \multicolumn{3}{c}{Width $\Gamma$ [$cm^{-1}$]} \\
      J   & v  & Present,TDSE            & Present & Experiment~\cite{Jastrzebski2009} & $\Delta E$ & Present,TDSE           & Present & $\Delta \Gamma$ \\
      	\midrule              
      1   & 18 & 21140.338          & 21140.449              &            &          	   &1.2580                 &1.3698                 & \bigdif{0.089} \\
      2   & 18 & 21140.605          & 21140.720              &            &          	   &1.3096                 &1.4321                 & \bigdif{0.094} \\
      3   & 18 & 21141.006          & 21141.122              &            &          	   &1.3787                 &1.5282                 & \bigdif{0.108} \\
      4   & 18 & 21141.507          & 21141.643              &            &          	   &1.4948                 &1.6508                 & \bigdif{0.104} \\
      5   & 18 & 21142.175          & 21142.238              &            &          	   &1.6331                 &1.7715                 & \bigdif{0.085} \\
\HIDE{9   & 17 & 21123.106          & 21123.131              & 21123.635  &  0.529,  0.504 &1.0799$\times 10^{-9} $&9.4903$\times 10^{-10}$& \bigdif{-0.121}\\}
\HIDE{10  & 17 & 21124.942          & 21124.966              &            &          	   &2.0026$\times 10^{-8} $&2.1731$\times 10^{-8} $& \bigdif{0.085} \\}
\HIDE{11  & 17 & 21126.979          & 21126.973              & 21127.370  &  0.391,  0.397 &2.7966$\times 10^{-7} $&2.4276$\times 10^{-7} $& \bigdif{-0.132}\\}
\HIDE{12  & 17 & 21129.150          & 21129.150              & 21129.508  &  0.358,  0.358 &1.8633$\times 10^{-6} $&1.8238$\times 10^{-6} $&         -0.021 \\}
\HIDE{13  & 17 & 21131.488          & 21131.494              & 21131.791  &  0.303,  0.297 &9.5909$\times 10^{-6} $&1.0543$\times 10^{-5} $& \bigdif{0.099} \\}
\HIDE{14  & 17 & 21133.993          & 21133.998              & 21134.211  &  0.218,  0.213 &4.8781$\times 10^{-5} $&5.0431$\times 10^{-5} $&         0.034 \\}%
\HIDE{15  & 17 & 21136.664          & 21136.659              & 21136.835  &  0.171,  0.176 &2.0220$\times 10^{-4} $&2.0800$\times 10^{-4} $&         0.029 \\}
\HIDE{16  & 17 & 21139.470          & 21139.469              & 21139.584  &  0.114,  0.115 &7.4449$\times 10^{-4} $&7.5907$\times 10^{-4} $&         0.020 \\}
      17  & 17 & 21142.408          & 21142.423              & 21142.417  &  0.009, -0.006 &2.4877$\times 10^{-3} $&2.4938$\times 10^{-3} $&         0.002 \\
      18  & 17 & 21145.514          & 21145.509              & 21145.434  & -0.080, -0.075 &7.3878$\times 10^{-3} $&7.4499$\times 10^{-3} $&         0.008 \\
      19  & 17 & 21148.720          & 21148.719              & 21148.553  & -0.167, -0.166 &2.0125$\times 10^{-2} $&2.0341$\times 10^{-2} $&         0.011 \\
      20  & 17 & 21152.026          & 21152.037              & 21151.714  & -0.312, -0.323 &4.9872$\times 10^{-2} $&5.0865$\times 10^{-2} $&         0.020 \\
      21  & 17 & 21155.433          & 21155.451              & 21155.066  & -0.367, -0.385 &1.1437$\times 10^{-1} $&1.1653$\times 10^{-1} $&         0.019 \\
      22  & 17 & 21158.939          & 21158.948              & 21158.400  & -0.539, -0.548 &2.4177$\times 10^{-1} $&2.4461$\times 10^{-1} $&         0.012 \\
      23  & 17 & 21162.513          & 21162.521              & 21161.830  & -0.683, -0.691 &4.7034$\times 10^{-1} $&4.7253$\times 10^{-1} $&         0.005 \\
      24  & 17 & 21166.153          & 21166.174              &            &          	   &8.3086$\times 10^{-1} $&8.4896$\times 10^{-1} $&         0.022 \\
      25  & 17 & 21169.860          & 21169.917              &            &          	   &1.3596                 &1.4341                 &         0.055 \\
      26  & 17 & 21173.594          & 21173.538              &            &          	   &2.0983                 &2.1353                 &         0.018 \\
\HIDE{24  & 16 & 21135.729          & 21135.742              & 21135.324  & -0.405, -0.418 &3.6174$\times 10^{-10}$&3.9474$\times 10^{-10}$& \bigdif{0.091} \\}
\HIDE{25  & 16 & 21140.672          & 21140.685              & 21140.004  & -0.668, -0.681 &1.7337$\times 10^{-8} $&1.8265$\times 10^{-8} $&         0.054 \\}
\HIDE{26  & 16 & 21145.748          & 21145.778              & 21146.681  &  0.933,  0.903 &3.6235$\times 10^{-7} $&3.8485$\times 10^{-7} $&         0.062 \\}
\HIDE{27  & 16 & 21150.991          & 21151.012              & 21151.614  &  0.623,  0.602 &4.6940$\times 10^{-6} $&4.8716$\times 10^{-6} $&         0.038 \\}
\HIDE{28  & 16 & 21156.368          & 21156.380              & 21156.762  &  0.394,  0.382 &4.2461$\times 10^{-5} $&4.2963$\times 10^{-5} $&         0.012 \\}%
\HIDE{29  & 16 & 21161.845          & 21161.868              & 21162.014  &  0.169,  0.146 &2.8348$\times 10^{-4} $&2.8770$\times 10^{-4} $&         0.015 \\}
      30  & 16 & 21167.455          & 21167.465              & 21167.380  & -0.075, -0.085 &1.5023$\times 10^{-3} $&1.5443$\times 10^{-3} $&         0.028 \\
      31  & 16 & 21173.133          & 21173.152              & 21172.888  & -0.245, -0.264 &6.7315$\times 10^{-3} $&6.8733$\times 10^{-3} $&         0.021 \\
      32  & 16 & 21178.877          & 21178.906              & 21178.406  & -0.471, -0.500 &2.5554$\times 10^{-2} $&2.5844$\times 10^{-2} $&         0.011 \\
      33  & 16 & 21184.688          & 21184.702              & 21183.965  & -0.723, -0.737 &8.1813$\times 10^{-2} $&8.2820$\times 10^{-2} $&         0.012 \\
      34  & 16 & 21190.499          & 21190.509              & 21189.630  & -0.869, -0.879 &2.2249$\times 10^{-1} $&2.2701$\times 10^{-1} $&         0.020 \\
      35  & 16 & 21196.309          & 21196.311              &            &          	   &5.2662$\times 10^{-1} $&5.3651$\times 10^{-1} $&         0.019 \\
      36  & 16 & 21202.087          & 21202.120              &            &          	   &1.0863                 &1.1147                 &         0.026 \\
      37  & 16 & 21207.864          & 21207.938              &            &          	   &1.9701                 &2.0734                 &         0.052 \\
\HIDE{34  & 15 & 21154.164          & 21154.168              & 21153.802  & -0.362, -0.366 &3.3995$\times 10^{-10}$&3.1517$\times 10^{-10}$&        -0.073 \\}
\HIDE{35  & 15 & 21161.478          & 21161.472              & 21162.212  &  0.734,  0.740 &1.2180$\times 10^{-8} $&1.2665$\times 10^{-8} $&         0.040 \\}
\HIDE{36  & 15 & 21168.891          & 21168.907              &            &          	   &2.7307$\times 10^{-7} $&2.7588$\times 10^{-7} $&         0.010 \\}
\HIDE{37  & 15 & 21176.472          & 21176.463              & 21176.701  &  0.229,  0.238 &3.8768$\times 10^{-6} $&3.8614$\times 10^{-6} $&        -0.004 \\}
\HIDE{38  & 15 & 21184.120          & 21184.126              & 21184.189  &  0.069,  0.063 &3.7771$\times 10^{-5} $&3.8557$\times 10^{-5} $&         0.021 \\}
\HIDE{39  & 15 & 21191.868          & 21191.883              & 21191.743  & -0.125, -0.140 &2.8978$\times 10^{-4} $&2.9404$\times 10^{-4} $&         0.015 \\}
      40  & 15 & 21199.716          & 21199.713              & 21199.333  & -0.383, -0.380 &1.7736$\times 10^{-3} $&1.7881$\times 10^{-3} $&         0.008 \\
      41  & 15 & 21207.564          & 21207.593              & 21206.890  & -0.674, -0.703 &8.7548$\times 10^{-3} $&8.9113$\times 10^{-3} $&         0.018 \\
      42  & 15 & 21215.479          & 21215.488              &            &          	   &3.6380$\times 10^{-2} $&3.6912$\times 10^{-2} $&         0.015 \\
      43  & 15 & 21223.327          & 21223.358              &            &          	   &1.2539$\times 10^{-1} $&1.2745$\times 10^{-1} $&         0.016 \\
      44  & 15 & 21231.141          & 21231.163              &            &          	   &3.6025$\times 10^{-1} $&3.6692$\times 10^{-1} $&         0.019 \\
      45  & 15 & 21238.856          & 21238.891              &            &          	   &8.7055$\times 10^{-1} $&8.8924$\times 10^{-1} $&         0.021 \\
      46  & 15 & 21246.504          & 21246.573              &            &          	   &1.7858                 &1.8662                 &         0.045 \\
\HIDE{40  & 14 & ???                & 21155.133              & 21156.143  &  ???,    1.010 & ???                   &0.1637$\times 10^{-14}$&         ???   \\}
\HIDE{41  & 14 & ???                & 21164.261              & 21165.066  &  ???,    0.805 & ???                   &0.5169$\times 10^{-12}$&         ???   \\}
\HIDE{42  & 14 & 21173.500          & 21173.529              &            &          	   &3.9293$\times 10^{-10}$&5.0401$\times 10^{-11}$&\bigdif{-0.872} \\}
\HIDE{43  & 14 & 21182.918          & 21182.928              &            &          	   &2.3312$\times 10^{-9} $&2.2425$\times 10^{-9} $&        -0.038 \\}
\HIDE{44  & 14 & 21192.436          & 21192.447              &            &          	   &5.6469$\times 10^{-8} $&5.6882$\times 10^{-8} $&         0.007 \\}
\HIDE{45  & 14 & 21202.054          & 21202.075              &            &          	   &9.1561$\times 10^{-7} $&9.4413$\times 10^{-7} $&         0.031 \\}
\HIDE{46  & 14 & 21211.772          & 21211.797              &            &          	   &1.1082$\times 10^{-5} $&1.1222$\times 10^{-5} $&         0.013 \\}
\HIDE{47  & 14 & 21221.590          & 21221.598              &            &          	   &1.0045$\times 10^{-4} $&1.0147$\times 10^{-4} $&         0.010 \\}
\HIDE{48  & 14 & 21231.442          & 21231.456              &            &          	   &7.1820$\times 10^{-4} $&7.2757$\times 10^{-4} $&         0.013 \\}
      49  & 14 & 21241.327          & 21241.345              &            &          	   &4.2147$\times 10^{-3} $&4.2525$\times 10^{-3} $&         0.009 \\
      50  & 14 & 21251.246          & 21251.231              &            &          	   &2.0288$\times 10^{-2} $&2.0543$\times 10^{-2} $&         0.013 \\
      51  & 14 & 21261.064          & 21261.066              &            &          	   &8.1186$\times 10^{-2} $&8.2309$\times 10^{-2} $&         0.014 \\
      52  & 14 & 21270.782          & 21270.793              &            &          	   &2.6828$\times 10^{-1} $&2.7182$\times 10^{-1} $&         0.013 \\
      53  & 14 & 21280.367          & 21280.376              &            &          	   &7.3356$\times 10^{-1} $&7.4032$\times 10^{-1} $&         0.009 \\
      54  & 14 & 21289.785          & 21289.840              &            &          	   &1.6714                 &1.6977                 &         0.016 \\
\HIDE{50  & 13 & 21205.527          & 21205.561              &            &          	   &6.4856$\times 10^{-9} $&2.5553$\times 10^{-10}$&\bigdif{-0.961} \\}
\HIDE{51  & 13 & 21217.015          & 21217.045              &            &          	   &9.2430$\times 10^{-9} $&7.5750$\times 10^{-9} $&\bigdif{-0.180} \\}
\HIDE{52  & 13 & 21228.603          & 21228.631              &            &          	   &1.4530$\times 10^{-7} $&1.4863$\times 10^{-7} $&         0.023 \\}
\HIDE{53  & 13 & 21240.259          & 21240.305              & 21240.344  &  0.085,  0.039 &2.0540$\times 10^{-6} $&2.0915$\times 10^{-6} $&         0.018 \\}
\HIDE{54  & 13 & 21252.014          & 21252.050              &            &          	   &2.1845$\times 10^{-5} $&2.2362$\times 10^{-5} $&         0.024 \\}
\HIDE{55  & 13 & 21263.803          & 21263.846              &            &          	   &1.8477$\times 10^{-4} $&1.8892$\times 10^{-4} $&         0.022 \\}
      56  & 13 & 21275.625          & 21275.668              &            &          	   &1.2668$\times 10^{-3} $&1.2967$\times 10^{-3} $&         0.024 \\
      57  & 13 & 21287.447          & 21287.484              &            &          	   &7.1898$\times 10^{-3} $&7.3604$\times 10^{-3} $&         0.024 \\
      58  & 13 & 21299.236          & 21299.248              &            &          	   &3.3983$\times 10^{-2} $&3.4759$\times 10^{-2} $&         0.023 \\
      59  & 13 & 21310.891          & 21310.900              & 21309.310  & -1.581, -1.590 &1.3478$\times 10^{-1} $&1.3582$\times 10^{-1} $&         0.008 \\
      60  & 13 & 21322.379          & 21322.373              &            &          	   &4.3449$\times 10^{-1} $&4.3478$\times 10^{-1} $&         0.001 \\
      61  & 13 & 21333.634          & 21333.644              &            &          	   &1.1463                 &1.1435                 &        -0.002 \\
      62  & 13 & 21344.721          & 21344.746              &            &          	   &2.5271                 &2.5240                 &        -0.001 \\
      \bottomrule
\end{tabular*}
\end{table*}

\subsection{Photodissociation dynamics}

\begin{figure}
\centering
\includegraphics[height=9cm,angle=270]{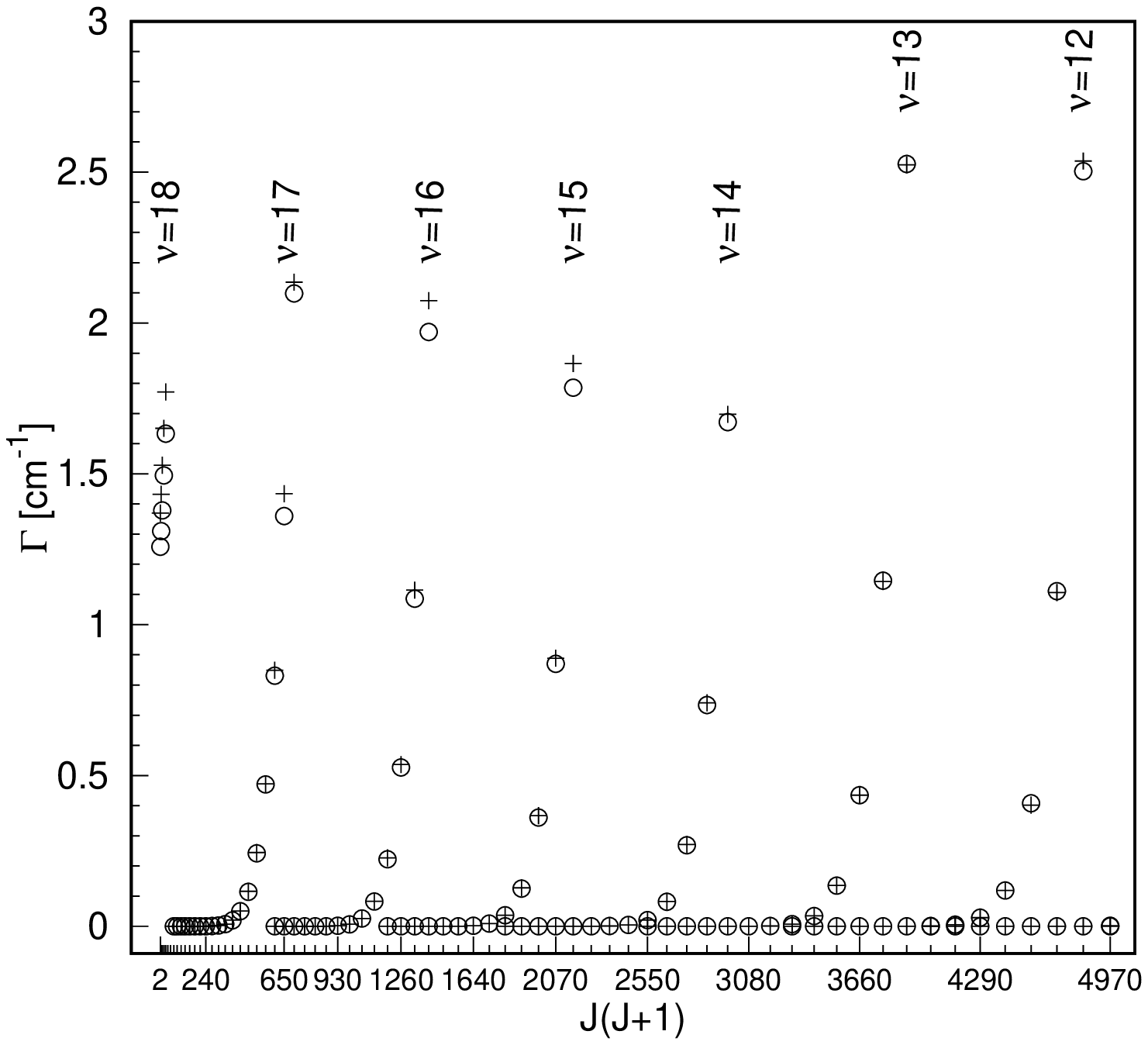}
\caption{Comparison of the widths of the bound ($\Gamma =0$) and quasi-bound levels. Our results from bound-continuum time-dependent simulations are given by the open circles, while results obtained from the program LEVEL are given by crosses~\cite{LeRoy2011}. 
}
\label{fig:level_widths}
\end{figure}

Using the computer code solving the time-dependent Schr\"odinger equation~\cite{Kosloff1984,Kosloff1997,Kosloff1988,Kozicki2008,Kozicki2009}, we are able to calculate several interesting features of the spectra and where it is possible to compare them with the other theoretical results and experimental data. In Fig.~\ref{fig:level_widths} we compare our results of the level widths $\Gamma$ (e.g. the full width at half maximum, FWHM) with those calculated from the program LEVEL~\cite{LeRoy2011}. Our line widths are calculated directly from the exponential decay of population. The extracted decay constant is used to estimate the width $\Gamma=\hbar/\tau$.  The program LEVEL calculates the line width from a uniform semi-classical treatment.
It is observed that the semi-classical treatment overestimates the line widths for the broadest levels.
Thus, the differences between the results become significant when the energy of quasi-bound levels approaches the top of the barrier. 
It is known that in this energy regime, the semi-classical approach does give less reliable results~\cite{LeRoy2011}. 

\begin{figure}
\centering
\includegraphics[height=\columnwidth,angle=270]{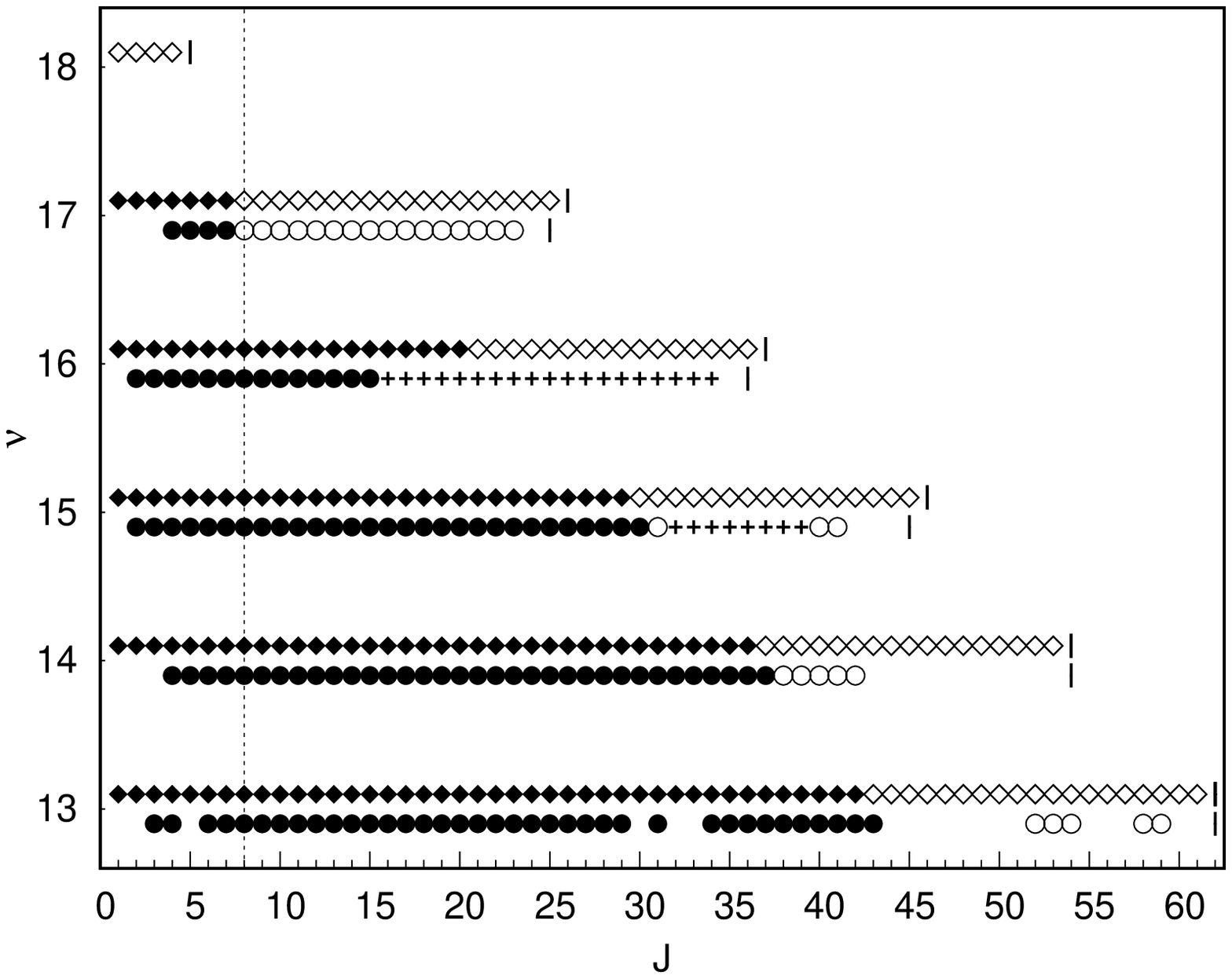}
\caption{Distribution of the present theoretical results (diamonds in the upper rows) and experimental data \cite{Jastrzebski2009} (circles in the lower rows) for the high rovibrational
levels in the $2^1\Pi$ state of KLi. Solid figures represent bound levels, open figures - quasi-bound levels, crosses - perturbed levels. Short vertical bars denote the last rotational level for each $v$.\\ The long broken vertical line placed at $J=8$ cuts $v$ levels involved in transitions showed in Fig. 7 . 
}

\label{fig:bound_quasi_bound}
\end{figure}

Our distribution of bound and quasi-bound levels for high vibrational numbers $v$ exhibits several similarities to the experimental data (Fig.~\ref{fig:bound_quasi_bound}). For $v=$13, 14 and 15 the values of $J$ which constitutes a boundary between bound and quasi-bound levels differs only by one. For $v=$16, because of strong perturbation with other states the comparison is impossible, while for $v=17$, there is a perfect agreement. At last for $v$=18, no experimental data is available. The long broken vertical line placed at $J=8$ crosses bound level with $v = 16$ and quasi-bound level with $v=17$, and clearly illustrating that the last line with $v=18$ in Fig. 7 comes from the virtual state lying above the potential barrier.   

\begin{figure}
\centering
\includegraphics[width=\columnwidth]{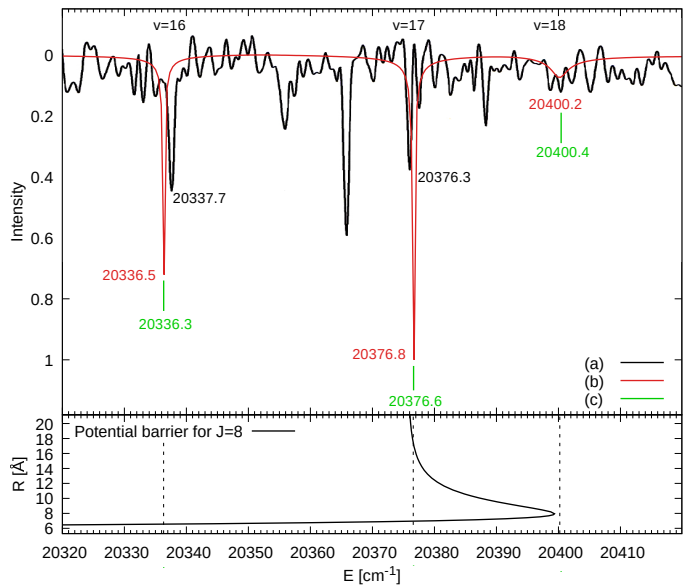}
\caption{The black color gives to the experimental spectrum of KLi observed with a linearly polarized pump laser beam \cite{Grochola2003, Grochola2017}. The assigned progression corresponds to  transitions from the ground state levels $v"=3$, $J"=8$ to the levels $v'=16$, $17$, and $18$, for $J'=8$ of $2^1\Pi$. The red color gives our spectrum calculated from Eq.\ref{tacs} using our potential curve of $2^1\Pi$. The green color is the spectrum calculated from the program LEVEL.
}
\label{fig:experiment_Grochola}
\end{figure}

The part of our discrete spectrum calculated from Eq.~\ref{tacs} for the absorption transitions from the ground state level $v"=3$, $J"=8$ to the levels $v'=16-18$, $J'=8$ of $2^1\Pi$ is compared with the experimental spectrum in Fig.~\ref{fig:experiment_Grochola}. According to our knowledge the remaining peaks in the spectrum are as yet not assigned by experimentalists. 
It should be noted that we have used the Condon approximation, i.e. a constant transition-dipole moment
and the relative intensities of the peaks may change when this approximation is abandoned. 
{The calculated spectrum is shifted by 744.835~cm$^{-1}$ with respect to the data in Tab.~\ref{tbl:comparison_all}.} 
{This shift was calculated in the following manner: the experimental energy 726.485~cm$^{-1}$ ~\cite{Bednarska1998} of the level  $v"=3$, $J"=0$ in the ground state was incremented by 18.350~cm$^{-1}$} which is the energy difference between  the ground state levels $v"=3$, $J"=8$ and $v"=3$, $J"=0$. The agreement between our and experimental spectra is very satisfactory showing the usability of our methods in assigning particular transitions to quasi-bound levels.  \JES{The process of quantum tunneling and forming the spectral lines can be viewed in the molecular movies (see Supplementary material~\cite{Supplementary2018}).}

\begin{figure}
\centering
\includegraphics[height=\columnwidth,angle=270]{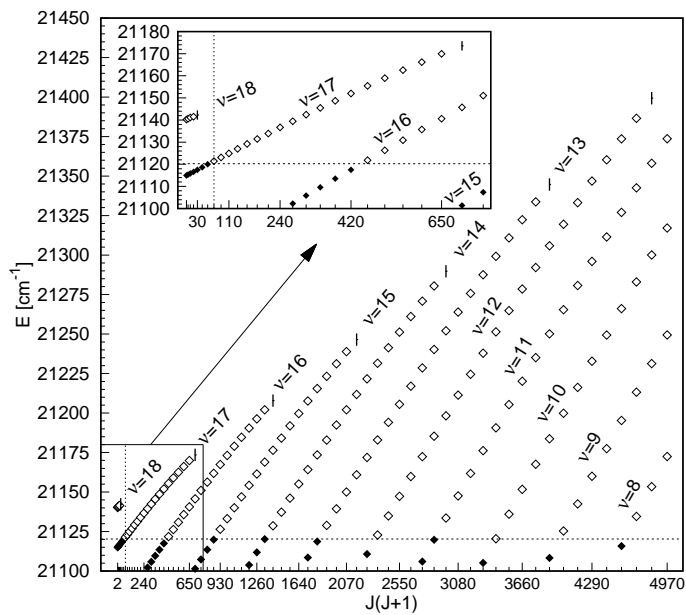}
\caption{Energies of term values for $v$ = 8-18 obtained from time-dependent calculations. Full diamonds indicate bound levels, empty diamonds - quasi-bound levels, short vertical bars - the last rotational level for a chosen $v$. 
}
\label{fig:all_terms}
\end{figure}

An overview of the calculated term energies may be found in Fig.~\ref{fig:all_terms}. The present results of term values from $v=8$ up to $v=18$ show a characteristic pattern. The numerical values of level energies and widths of the $2^1\Pi$ state may be found in Tab. 3. Here only levels with widths broader then 10$^{-3}$ cm$^{-1}$ are presented. As expected the biggest differences between theoretical results occur for the quasi-bound states lying just below the barrier's maximum, where the semi-classical  approach becomes insufficient. The differences between theoretical results and available experimental data do not exceed the experimental resolution. It means that these set of results may be useful in further assigning of experimental spectrum.  The energies, widths \JES{and lifetimes} of all calculated quasi-bound levels may be found in Supplementary material~\cite{Supplementary2018}.

\quad
\section{Conclusions}

In order to describe the rovibrational predissociation  process of the KLi molecule we  calculated the low lying adiabatic potential energy curves with particular emphasis on the $2^1\Pi$ state. Our spectroscopic parameters are in a very good agreement with experimental values.
A small barrier to dissociation ($\sim 20\ {\rm cm^{-1}}$ for $J=0$) is identified.
Having the potential curve of $2^1\Pi$ state we calculate the rovibrational levels. The differences between the successive levels are compared with those derived from experimental data. The agreement again is very good, which means that the shape of the excited electronic state $2^1\Pi$ is reliable. 

Using the complementary time-dependent approach we solve the time-dependent nuclear Schr\"{o}dinger equation. The solutions show the evolving wavepacket originally placed on the effective potential curve. 
The spectrum is calculated as a Fourier transform of the autocorrelation function. The
differences between  successive peaks in the spectrum are compared with the experimental spectrum. 
In our calculations of the time-dependent population of the rovibrational $(v,J)$ levels, we focus on
initial wavepackets chosen as eigenfunctions of  quasi-bound states calculated with classical turning points.  This approach allows for the exact description of the rovibrational predissociation mechanism of the KLi molecule. We show explicitly that the population of a quasi-bound state decays exponentially in time. This approach can be easily used for other diatomic and even polyatomic molecules. 

We also describe the detailed procedure of calculating widths of quasi-bound vibrational levels \JES{with the high accuracy}. \JES{The present method for solving TDSE is often used as a benchmark for testing other numerical methods~\cite{1991Leforestier}.} \JES{It is especially important in the view of ultracold experiments as quasi-bond could possibly be explored for new ways of cooling molecules. states }  Extensive tables with calculated level energies, widths \JES{and lifetimes are} presented. \JES{Those results are of considerable relevance to the design of experiments and the development of approximate computational methods.} For available experimental data the comparison with our results gives very good agreement. Certainly, we hope that the results may be helpful in assigning transitions to quasi-bound levels. 

\quad
\section*{Acknowledgments}
This work was partially supported by the COST action XLIC (CM1204) of the
European Community and the Polish Ministry of Science and Higher Education. Calculations have been carried out using resources of the Academic Computer Centre in Gda\'nsk.

\bibliographystyle{rsc} 
 \bibliography{KLi_bibliography}
\end{document}